\documentclass[11pt]{article}

\usepackage[final]{acl}

\usepackage{times}
\usepackage{latexsym}

\usepackage[T1]{fontenc}
\usepackage[utf8]{inputenc}

\usepackage{microtype}

\usepackage{inconsolata}

\usepackage{graphicx}

\title{\textsc{CiteCheck}: Retrieval-Grounded Detection of LLM Citation Hallucinations in Scientific Text}

\usepackage{booktabs}
\usepackage{subcaption}
\usepackage{caption}
\usepackage[table,dvipsnames]{xcolor}
\usepackage{array}
\usepackage{multirow}
\usepackage{makecell}
\usepackage{float}
\usepackage{placeins}
\usepackage{fancyvrb}
\usepackage{listings}
\usepackage{threeparttable}
\usepackage{tabularx}
\usepackage{amsmath}
\usepackage{amssymb}
\usepackage{mathtools}
\usepackage{amsthm}
\usepackage{fontawesome5} % add in preamble
\usepackage{url}
\usepackage{tikz}
\usetikzlibrary{arrows.meta, positioning, decorations.pathmorphing, calc, fit}

\usepackage[most]{tcolorbox}
\tcbuselibrary{listings,breakable}

\newtcblisting{promptbox}{
  breakable,
  listing only,
  colback=black!3,
  colframe=black!30,
  boxrule=0.4pt,
  arc=1pt,
  left=4pt,
  right=4pt,
  top=4pt,
  bottom=4pt,
  width=\columnwidth,
  listing options={
    basicstyle=\scriptsize\ttfamily,
    breaklines=true,
    breakatwhitespace=false,
    columns=fullflexible,
    keepspaces=true,
    showstringspaces=false,
    tabsize=2
  },
  before skip=4pt,
  after skip=4pt
}
\graphicspath{{Image/}{Image/hallucination/}}
\newcolumntype{Y}{>{\centering\arraybackslash}X}
\newcommand{\decrease}[1]{\textcolor{red}{\scriptsize$\downarrow\,#1$}}
\newcommand{\citepCheck}{\textsc{CiteCheck}}

\author{  
\textbf{Khashayar Khajavi\textsuperscript{1,2,*}} \quad
  \textbf{Shaghayegh Sadeghi\textsuperscript{1,*}} \quad
  \textbf{Rise Adhikari\textsuperscript{1}} \quad
  \textbf{Alexander Tessier\textsuperscript{1}}
  \\
  \textsuperscript{1}FirstPrinciples Inc. \\
  \textsuperscript{2}School of Computing Science, Simon Fraser University \\
  \textsuperscript{*}Equal contribution \\
  \small{\textbf{Correspondence:}
  \href{mailto:connect@firstprinciples.com}{connect@firstprinciples.com}}
\\[0.5em]
\href{https://github.com/FirstPrinciplesLabs/citecheck}
{\scalebox{1.5}{\faGithub}\ \textbf{Code}}
}
\begin{document}
\maketitle

\vspace{-0.5em}

\begin{abstract}
Large language models (LLMs) are increasingly used to generate scientific reports, but they can produce references that appear plausible while containing corrupted metadata or pointing to papers that do not exist. We introduce \citepCheck{}, a hybrid framework for citation hallucination detection that verifies whether a citation corresponds to a real scholarly work and whether its metadata is faithful to that work. \citepCheck{} retrieves candidate publications from external scholarly sources, compares the citation against the retrieved candidate using a structured LLM verifier, and maps verifier scores into three labels: \textsc{Exact}, \textsc{Minor}, and \textsc{Major}. We also construct a 982-citation physics benchmark with controlled corruptions that capture both subtle metadata drift and fully fabricated references. On the held-out test set, \citepCheck{} achieves 88.7 macro-F1 and 88.9\% accuracy, outperforming GPT, Claude, and Gemini baselines, including web-search and few-shot variants. These results show that reliable citation verification benefits from combining scholarly retrieval, structured LLM-based comparison, and calibrated decision rules. 
\end{abstract}

\section{Introduction}

% Large language models are increasingly used to draft literature reviews and scientific summaries. The quality of such text depends not only on fluency, but on whether each cited source actually exists and is faithfully described. Models routinely produce citations that look plausible while changing the title, inventing a DOI, assigning the wrong year, or fabricating an entire paper \citep{misra2026detecting,abbonato2026checkifexist,press2024citeme}.

Large language models (LLMs) are rapidly becoming embedded in scientific workflows, from literature discovery and summarization to hypothesis generation, manuscript drafting, and report generation \citep{zhang2025exploring,luo2025llm4sr}. Furthermore, recent studies have documented their growing footprint in scientific writing itself, including measurable linguistic shifts in published papers and broader use across the research lifecycle \citep{liang2024mapping,kobak2025delving}. This shift creates new opportunities for accelerating research, but it also changes the reliability requirements of generated scientific text. In scholarly writing, factuality is not only a matter of producing fluent and plausible claims: each claim must be connected to a traceable literature, and each cited source must be recoverable and correctly described.

Citations are a core mechanism through which science maintains continuity, attribution, and accountability. They situate new work within prior knowledge, guide readers to supporting evidence, acknowledge intellectual contributions, and increasingly influence research evaluation and academic reputation \citep{gasparyan2015preserving,bruton2025citation,mehregan2022scientific}. Citation errors therefore do more than introduce formatting noise: they can mislead readers, obscure provenance, distort credit, and weaken the reproducibility of scientific communication. This problem becomes especially acute for LLM-generated content, where models can produce references that appear stylistically legitimate while changing authors, years, titles, DOIs, or URLs, or fabricating entire papers that do not exist \citep{walters2023fabrication,mcgowan2023chatgpt,linardon2025influence,misra2026detecting,abbonato2026checkifexist,press2024citeme}.

% We focus on citation existence and metadata fidelity. In practice, four failure modes recur: (i) \emph{metadata drift}, where a real paper is cited with incorrect surface fields; (ii) \emph{title fabrication}, where the citation points to a plausible work but the title is paraphrased or invented; (iii) \emph{whole-paper fabrication}, where no scholarly record exists; and (iv) \emph{citation swap}, where two real references are transposed. This paper addresses the first three as a three-class classification task; swaps are handled by a downstream alignment stage. The problem is closely related to broader work on attribution and evidence-grounded generation \citep{rashkin-etal-2023-measuring,gao-etal-2023-enabling,schreieder2025attribution}.

% We introduce \citepCheck{}, a hybrid system that parses each reference into structured metadata, searches for a real candidate through a waterfall cascade of scholarly sources, compares the candidate against the original citation with an LLM verifier that emits a 0--10 score, applies a deterministic threshold rule to assign exact, minor, or major, and rewrites recoverable minor hallucinations in canonical form.

In this work, we introduce \citepCheck{}, a hybrid framework for detecting hallucinated references in AI-generated scientific reports. Rather than relying on the surface plausibility of a citation, \citepCheck{} grounds each reference against external scholarly sources and uses a structured LLM verifier to judge whether the retrieved publication actually matches the cited metadata. The system distinguishes between three outcomes: \emph{exact matches}, where the citation correctly describes a real paper; \emph{minor hallucinations}, where the intended paper exists but some bibliographic fields are corrupted; and \emph{major hallucinations}, where the reference points to no matching scholarly work.

To evaluate this setting, we construct a benchmark of 982 real physics citations and controlled corrupted variants. The benchmark covers a diverse set of physics subfields and includes both subtle metadata perturbations, such as incorrect authors, years, titles, or identifiers, and fully fabricated references that remain topically plausible. On the test set, \citepCheck{} achieves 88.7 macro-F1 and 88.9\% accuracy, outperforming GPT, Claude, and Gemini baselines, including variants with web search and few-shot examples. Notably, \citepCheck{} remains zero-shot, yet surpasses the strongest few-shot LLM baseline by 5.8 macro-F1 points and 5.7 accuracy points, showing that grounded retrieval and structured comparison are more reliable than prompting LLMs to judge citation validity directly.

The rest of the paper is organized as follows. Section~\ref{sec:related} reviews work on LLM-assisted scientific writing, citation hallucination, citation integrity, and claim-level verification. Section~\ref{sec:method} presents the \citepCheck{} pipeline, including candidate retrieval, LLM-based verification, threshold-based labeling, and the reviewer pass. Section~\ref{sec:dataset} describes the construction of our citation-hallucination benchmark. Section~\ref{sec:results} reports the main comparison against LLM baselines, verifier-model sensitivity, and component ablations. Section~\ref{sec:discussion} discusses the implications of the approach, and Section \ref{sec:limitations} mentions the remaining limitations.
\begin{figure*}[t]
    \centering
    \includegraphics[width=1\textwidth]{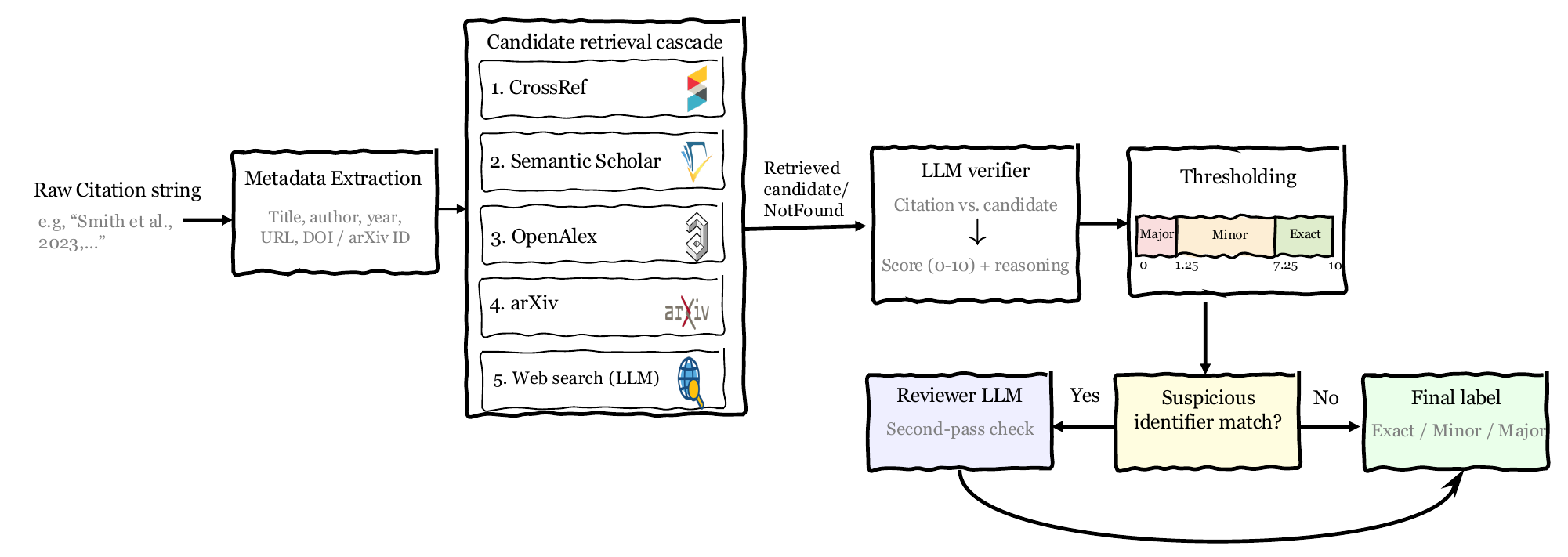}
    \caption{The \citepCheck{} pipeline. A raw citation string is first parsed into structured metadata, then matched against candidates from a retrieval cascade. An LLM verifier scores the citation against the retrieved candidate and the score is mapped to one of three labels (Exact / Minor / Major). When the candidate's identifier match is suspicious, a Reviewer LLM performs a second-pass check before the final label is emitted.}
    \label{fig:framework}
\end{figure*}

\section{Related Work}
\label{sec:related}

\paragraph{LLMs in scientific writing and citation hallucination.}
LLMs are increasingly used across scientific workflows, including literature review, summarization, hypothesis generation, and manuscript drafting \citep{zhang2025exploring,luo2025llm4sr}, and their footprint is now visible in the literature itself through measurable linguistic shifts in published papers \citep{liang2024mapping,kobak2025delving}. This trend motivates tools that verify not only fluency but also the integrity of the references such text depends on. A growing body of work shows that LLMs fabricate or distort bibliographic references, producing citations that appear plausible but contain invented titles, non-existent DOIs, incorrect authors, or fully fabricated papers \citep{walters2023fabrication,mcgowan2023chatgpt,linardon2025influence,misra2026detecting,abbonato2026checkifexist,press2024citeme}. Closest to our setting, \citet{abbonato2026checkifexist} introduce a tool for checking reference existence against scholarly metadata sources. \citepCheck{} differs by framing citation hallucination as a three-class severity task, distinguishing exact matches, minor metadata corruptions, and major fabricated references, and by combining scholarly retrieval, structured LLM scoring, threshold-based classification, and a reviewer pass in a single evaluation-driven pipeline.

\paragraph{Citation integrity and attribution.}
Citations are central to scholarly communication, connecting new work to prior evidence and supporting the traceability and reproducibility of research; inaccurate or manipulated references can mislead readers, distort credit, and weaken the reliability of the scholarly record \citep{gasparyan2015preserving,bruton2025citation,mehregan2022scientific}. A related line of work studies whether generated claims are properly supported by their cited sources \citep{gao-etal-2023-enabling,rashkin-etal-2023-measuring,huang2024learning,wu2025automated,zhang2024finegrained,yue2305automatic}. Long-form factuality metrics such as \textsc{FActScore} \citep{min-etal-2023-factscore}, self-reflective retrieval methods \citep{bohnet2022attributedqa,asai2024selfrag}, and scientific claim verification benchmarks \citep{wadden2020scifact,wadden2022scifactopen,liu2024cliver,alvarez2024zero,fang2025automatic,wang2025sciver,ho2026sciclaimeval} generally assume that references resolve to real documents. \citepCheck{} sits one level earlier: it verifies that cited works themselves exist and are correctly specified, complementing rather than replacing claim-level verification, and directly addresses citation-integrity concerns in AI-generated reports, where reference errors can be produced at scale and are difficult to detect manually.

\paragraph{Scholarly retrieval.}
\citepCheck{} relies on scholarly metadata services that are standard infrastructure for scientific search and reference resolution \citep{hendricks2020crossref,kinney2023semantic,priem2022openalex,arxiv_api}. Its retrieval component is related to retrieval-augmented generation \citep{lewis2020rag,karpukhin2020dense}, but the goal differs: rather than retrieving passages to support generation, \citepCheck{} retrieves candidate publications to validate citation metadata and determine whether an apparent reference corresponds to a real scholarly work.
\section{Method}
\label{sec:method}
% \subsection{Problem Formulation}
% Let $R=\{r_1,\dots,r_n\}$ be the references parsed from a report, with each
% $r_i=(a_i, y_i, t_i, u_i, x_i)$ giving authors, year, title, URL, and arXiv identifier. Let $\mathcal{P}$ be the universe of real publications and $\phi:R\to\mathcal{P}\cup\{\bot\}$ the latent mapping from a reference to its true publication ($\bot$ if none exists). The goal is to assign each reference a label
% $\ell_i\in\{\textsc{Exact},\textsc{Minor},\textsc{Major}\}$:
% \textsc{Exact} when all observed fields are consistent with the matched publication;
% \textsc{Minor} when a real publication exists but at least one field is incorrect;
% \textsc{Major} when no matching publication exists or the closest candidate is too dissimilar to be the same work.
% For minor hallucinations, the correction task is to render a canonical citation string from the matched metadata.
\subsection{Problem Formulation}

We study citation hallucination detection at the reference level. Given a citation string, the goal is to determine whether it corresponds to a real scholarly work and whether its bibliographic metadata is faithful to that work. We focus on citation existence and metadata fidelity, rather than whether the cited paper supports a particular claim.

We formulate the task as three-way classification. An \textsc{Exact} citation correctly identifies a real publication and contains metadata consistent with that publication. A \textsc{Minor} hallucination refers to a real publication, but contains one or more incorrect fields, such as an altered author name, year, title, URL, DOI, or arXiv identifier. A \textsc{Major} hallucination is a citation for which no sufficiently similar real publication can be found, or where the closest retrieved candidate is too different to plausibly be the intended work.

\subsection{Overview}
\label{sec:method-overview}

Figure~\ref{fig:framework} gives an overview of \citepCheck{}. The input is a single citation string. \citepCheck{} first converts this raw citation into structured fields, including title, authors, year, URL, DOI, and arXiv identifier when available. We use a rule-based parser for common citation patterns and invoke an LLM-based parser only when the deterministic parser fails to recover the core fields. This step does not determine whether the citation is valid; it only extracts retrieval signals for the later stages.

The central idea of \citepCheck{} is to separate \emph{retrieval} from \emph{verification}. A citation can look plausible even when it is wrong, and a real identifier can be paired with an unrelated title or author list. Rather than asking an LLM to judge citation validity from its internal knowledge, \citepCheck{} first attempts to ground the citation in external scholarly sources. It searches for the closest real publication using a waterfall cascade over scholarly metadata APIs and web search. The output of this stage is either the best candidate publication found for the citation or a not-found result when no reliable candidate can be recovered.

Given the original citation and the retrieved candidate, \citepCheck{} performs structured verification. A verifier LLM compares the citation against the candidate metadata and assigns a numeric score from 0 to 10, together with a short explanation of the main discrepancies. The score is then mapped to one of three labels using fixed thresholds: \textsc{Exact}, \textsc{Minor}, or \textsc{Major}. High scores indicate that the citation faithfully describes the retrieved work, intermediate scores indicate that the intended work likely exists but contains corrupted metadata, and low scores indicate that the citation is fabricated or too dissimilar from any retrieved publication. For suspicious identifier-based matches, i.e., where a URL or arXiv ID resolves to a real paper but the title similarity is low, \citepCheck{} applies an optional reviewer pass that re-examines the case before assigning the final label.

\subsection{Retrieving Candidate Publications}
\label{sec:candidate-retrieval}

Given an input citation, \citepCheck{} first attempts to recover the real publication that the citation most likely refers to. This step is necessary because hallucinated citations are often only partially wrong: they may preserve the correct title while corrupting the year, URL, DOI, or arXiv identifier, or pair a real identifier with unrelated surrounding metadata. We therefore treat retrieval as candidate generation rather than final verification.

\citepCheck{} constructs a cleaned search query from the parsed title when available, falling back to the raw citation text otherwise. Common LaTeX and BibTeX artifacts are removed before querying, and the publication year, when available, is used as an auxiliary ranking signal. The system then applies a waterfall cascade over external sources. If the citation appears arXiv-related, a direct arXiv lookup can be attempted first. The cascade then searches CrossRef~\citep{crossref_api}, Semantic Scholar~\citep{semanticscholar_api}, and OpenAlex~\citep{priem2022openalex}. These sources provide complementary coverage: CrossRef is strong for DOI-registered publisher metadata, Semantic Scholar provides broad scholarly search with enriched metadata, and OpenAlex offers a large open index of scholarly works. If no structured source returns a reliable match, \citepCheck{} optionally falls back to LLM-assisted web search.

At each stage, \citepCheck{} records the best candidate returned by that source, but accepts a candidate only if it passes a minimum title-similarity threshold. CrossRef additionally requires a confidence threshold, while later sources can replace the CrossRef candidate only when they achieve higher title similarity. Candidate acceptance is based on normalized Levenshtein similarity between cleaned titles. Within Semantic Scholar and OpenAlex, a cheaper word-overlap score with year bonuses is used only to rank returned candidates before the final title-similarity check. Additional implementation details for query construction, scoring, confidence handling, and fallback behavior are provided in Appendix~\ref{app:candidate-retrieval}.

The retrieval stage outputs either a single best candidate publication or a not-found result. Importantly, this output is not treated as the final answer: high title similarity can still hide metadata drift, and identifier-based matches can be misleading when a real URL or arXiv ID is attached to the wrong title or authors. The retrieved candidate is therefore passed to the LLM verification stage for direct comparison against the original citation. Furthermore, we provide an efficiency analysis of the retrieval cascade, including per-stage latency and match-source distributions, in Appendix~\ref{app:cascade-latency}.

\subsection{LLM-Based Verification and Label Assignment}
\label{sec:llm-verification}

After retrieval, \citepCheck{} verifies whether the input citation faithfully describes the retrieved candidate. The verifier receives both the original citation metadata and the matched publication metadata, and outputs a structured response containing a numeric score from 0 to 10, a confidence level, a short rationale, and the main differences it identifies. Importantly, the LLM is not asked to decide citation validity from memory; it is used as a comparator over externally retrieved evidence.

We use a numeric score rather than relying directly on the model's categorical label. This allows the final decision boundary between \textsc{Exact}, \textsc{Minor}, and \textsc{Major} to be tuned on validation data, instead of being fixed entirely by prompt wording. Similar uses of graded LLM outputs have been explored in settings where downstream thresholding, ranking, or calibration is useful, including Likert-style LLM screening for literature review and calibrated verbalized probabilities for discriminative tasks \citep{dennstadt2024title,xiong2024calibrating}. In our setting, the score provides a continuous measure of citation--candidate agreement: high scores correspond to faithful metadata matches, intermediate scores indicate a recognizable paper with corrupted fields, and low scores indicate an unrelated or fabricated citation.

The final label is assigned deterministically from the verifier score using two thresholds, $\tau_M$ and $\tau_E$:
\[
\ell =
\begin{cases}
\textsc{Exact}, & s \geq \tau_E, \\
\textsc{Minor}, & \tau_M \leq s < \tau_E, \\
\textsc{Major}, & s < \tau_M.
\end{cases}
\]
The thresholds are selected on the development split and then held fixed for test evaluation. This separation makes re-calibration inexpensive: once the verifier scores have been produced, different threshold choices can be evaluated without re-running retrieval or LLM inference. Prompt details and the structured output schema are provided in Appendix~\ref{app:llm-verification}.

\subsection{Reviewer Pass for Suspicious Matches}
\label{sec:reviewer-pass}

A remaining failure mode occurs when an identifier resolves to a real paper, but the visible citation text describes a different work. For example, a fabricated citation may contain a valid arXiv URL while pairing it with unrelated authors or a different title. To handle this, \citepCheck{} uses a targeted reviewer pass for suspicious cases where the retrieved candidate comes from an identifier-based source, such as arXiv or web search, but the title similarity is low.

The reviewer is implemented as a second LLM judge that receives the raw citation, the matched source metadata, the title-similarity score, and the first-pass verifier's label, score, and reasoning. This design follows the broader use of LLMs as structured evaluators, where a model is given explicit criteria and evidence to judge an earlier output \citep{zheng2023judging,liu2023geval}. The reviewer outputs only a final label and rationale; when it overrides the first-pass decision, \citepCheck{} maps that label back to a representative score within the corresponding threshold interval while preserving the original verifier output for auditability. The reviewer prompt is provided in Appendix~\ref{app:reviewer-prompt}.

\section{Dataset Construction}
\label{sec:dataset}

\subsection{Citation Pool}
\label{sec:citation-pool}

We construct our benchmark from a pool of 982 real academic citations drawn from physics literature. The pool was curated to reflect realistic citation contexts in scientific reports: most references are research papers, including preprints and peer-reviewed articles, while a small number correspond to other scholarly artifacts such as theses, technical reports, facility documentation, and canonical scientific databases. The citations span nine subdomains: astrophysics, biophysics, condensed matter, gravitational physics, nuclear physics, particle physics, plasma physics, quantum computing, and soft matter physics. To avoid a benchmark where topical mismatch becomes an easy shortcut, we organize the citations into 42 topically coherent collections. Within each collection, references share a common subfield and technical vocabulary, making the task depend on citation-level evidence rather than broad topic recognition.

Each citation in the pool corresponds to a real, retrievable scholarly work and serves as a valid reference before corruption. We normalize all citations into a common surface format containing the author string, publication year, paper title, and a resolvable URL, such as a DOI, arXiv link, or publisher page. This standardization allows the corruption and verification pipelines to operate over a consistent input representation while preserving the bibliographic fields most relevant to citation hallucination detection.

The pool is intentionally representative of modern physics citation practices. Most references resolve to arXiv, reflecting the central role of preprints in the selected subdomains, while the remaining citations come from publisher and indexing platforms. The publication years span both established and recent work, which is important because verification systems may rely on metadata sources with uneven coverage over time. Table~\ref{tab:dataset-subdomains} summarizes the distribution of citations across subdomains.

\begin{table}[t]
\centering
\small
\caption{Distribution of the original citation pool across physics subdomains.}
\label{tab:dataset-subdomains}

\begin{tabularx}{\columnwidth}{>{\raggedright\arraybackslash}Xrr}
\toprule
Subdomain & Collections & Citations \\
\midrule
Astrophysics & 4 & 92 \\
Biophysics & 3 & 76 \\
Condensed matter & 5 & 104 \\
Gravitational physics & 5 & 135 \\
Nuclear physics & 5 & 92 \\
Particle physics & 5 & 124 \\
Plasma physics & 5 & 121 \\
Quantum computing & 5 & 134 \\
Soft matter physics & 5 & 104 \\
\midrule
Total & 42 & 982 \\
\bottomrule
\end{tabularx}

\end{table}

\subsection{Controlled Corruption Procedure}
\label{sec:controlled-corruptions}
Starting from the clean citation pool, we create a labeled benchmark by applying controlled corruptions to a stratified subset of citations while leaving the rest unchanged. Within each topical collection, citations are randomly partitioned using a fixed seed into valid citations, minor hallucinations, and major hallucinations, so that all physics subdomains contain a similar label mix and the benchmark cannot be solved by topic-level artifacts.
Valid citations are kept unchanged. Minor hallucinations model metadata drift: the intended paper still exists, but one or more bibliographic fields, e.g., author names, year, URL, DOI, arXiv identifier, or venue, are corrupted, while the title is kept mostly intact. Major hallucinations model fabricated references: the citation is rewritten into a plausible but non-existent paper, conditioned on the original subdomain and with coordinated changes across title, authors, year, and identifier fields.
Both corruption types are generated with the same LLM backbone (GPT-4o-mini) under different temperature and prompting regimes, low temperature for conservative, transcription-error-like minor edits and high temperature for more substantial major fabrications, so that differences between classes arise from the corruption regime rather than generator identity. Each generation call returns a structured record with the original citation, corrupted citation, label, and a natural-language description of the changes, providing a provenance trace for every example. Full prompt templates, schema details, and examples are in Appendix~\ref{app:benchmark-construction}.

\subsection{Dataset Statistics}
\label{sec:dataset-statistics}

The final benchmark contains 982 citation instances. The label distribution is close to balanced across the three classes: 356 valid citations, 300 minor hallucinations, and 326 major hallucinations. The slight skew toward valid citations comes from the deterministic rounding rule used when partitioning citations within each topical collection. Table~\ref{tab:label-distribution} summarizes the final label distribution.

\begin{table}[t]
\centering
\small
\caption{Label distribution in the final benchmark.}
\label{tab:label-distribution}

\begin{tabularx}{\columnwidth}{>{\raggedright\arraybackslash}Xrr}
\toprule
Label & Count & Share \\
\midrule
Valid & 356 & 36.25\% \\
Minor hallucination & 300 & 30.55\% \\
Major hallucination & 326 & 33.20\% \\
\midrule
Total & 982 & 100.00\% \\
\bottomrule
\end{tabularx}

\end{table}

The benchmark is also diverse in source type and publication year. Since the corpus is drawn from physics, it is preprint-heavy: 835 citations, or 85.0\%, resolve to arXiv, while the remaining citations resolve to publisher pages, indexing services, or biomedical repositories. Publication years range from 2007 to 2025, with a median year of 2020. This mix is useful for evaluation because citation-verification systems must handle both recent preprints and older, well-established publications. All 982 references resolve to real scholarly artifacts. Of these, 942 are research papers, including arXiv preprints and peer-reviewed journal articles, while the remaining references consist of theses, technical reports, public-facility briefings, and canonical scientific databases.

Because corruptions are applied independently within each topical collection, the three labels are distributed similarly across physics subdomains. This design reduces the chance that a model can exploit topic-level artifacts to distinguish valid and hallucinated citations. Instead, successful systems must verify whether the citation metadata corresponds to a real publication.

\section{Results}
\label{sec:results}

% Table~\ref{tab:main_results} reports three-class classification on the hallucinated test set. \citepCheck{} achieves 86.1 exact $F_1$, 81.7 minor $F_1$, 98.3 major $F_1$, 88.7 macro $F_1$, and 88.9 accuracy, the best results across all metrics. The strongest baseline (few-shot Claude Sonnet 4.6 with web search) reaches 82.9 macro $F_1$ and 83.2 accuracy, leaving a 5.8-point macro gap. Retrieval, structured verification, deterministic thresholding, and the reviewer pass all contribute beyond what prompting alone provides.

% The ablation in Table~\ref{tab:ablation} isolates each contribution. Removing the reviewer pass costs 4.0 macro $F_1$, with the largest drop on minor hallucinations (--6.0). Removing the web search fallback costs 7.0 macro $F_1$, with the largest drop again on minor hallucinations (--9.0). Broader candidate retrieval matters most when citation metadata is heavily distorted.

\subsection{Experimental Setup}
\label{sec:experimental-setup}

We evaluate \citepCheck{} on the 982-citation benchmark described in Section~\ref{sec:dataset}. We split the benchmark into a validation set of 190 citations and a held-out test set of 792 citations. The validation set is used only for selecting the score thresholds that map verifier scores to the three output labels; all reported headline results are computed on the test set.

As mentioned in Section~\ref{sec:llm-verification}, \citepCheck{} produces a continuous score for each citation through its LLM verifier. To convert these scores into labels, we tune two thresholds on the validation set: a minor threshold $\tau_M$ and an exact threshold $\tau_E$, with the constraint $\tau_E \geq \tau_M$. We sweep both thresholds from 0 to 10 in increments of 0.25 and reclassify the validation predictions under each candidate pair. The selected threshold pair is the one that maximizes support-weighted F1 across the three classes. This procedure selects $\tau_M=1.25$ and $\tau_E=7.25$, which are then fixed and applied unchanged to the test set.

Unless otherwise stated, \citepCheck{} uses Claude Sonnet 4.6 as both the primary citation verifier and the reviewer model, and uses GPT-5.4 for the web-search fallback in the retrieval cascade. All LLM-based components are accessed through hosted provider APIs. We report per-class F1 for \textsc{Exact}, \textsc{Minor}, and \textsc{Major}, along with macro-F1 and overall accuracy. For LLM baselines, we evaluate GPT 5.4, Claude Sonnet 4.6, and Gemini 2.5 Flash under zero-shot and few-shot prompting, both with and without web search. The few-shot setting uses 27 examples sampled from the validation set, with nine examples from each class, and the same examples are used for all baseline models. For all direct LLM baselines, we use a shared citation-verification rubric that defines the same three labels as \citepCheck{} and requires a structured JSON output containing a label, score, reasoning, and whether evidence was found. The full baseline prompt templates are provided in Appendix~\ref{app:baseline-prompts}, and model/provider details, decoding settings, experiment dates, and tuned thresholds are reported in Appendix~\ref{app:llm-reproducibility}.

\subsection{Comparison with LLM Baselines}
\label{sec:baseline-comparison}

Table~\ref{tab:main_results} compares \citepCheck{} against direct LLM baselines on the held-out test set. \citepCheck{} achieves the best performance across all reported metrics, with 86.1 \textsc{Exact} F1, 81.7 \textsc{Minor} F1, 98.3 \textsc{Major} F1, 88.7 macro-F1, and 88.9\% accuracy. The strongest baseline is Claude Sonnet 4.6 with both web search and few-shot examples, which reaches 82.9 macro-F1 and 83.2\% accuracy. Thus, \citepCheck{} improves over the strongest baseline by 5.8 macro-F1 points and 5.7 accuracy points. Importantly, \citepCheck{} does not use any few-shot examples in its verifier prompt; it outperforms all zero-shot and few-shot baselines through retrieval-grounded verification and threshold calibration rather than in-context supervision. This baseline represents a direct LLM+web-search approach without \citepCheck{}'s scholarly retrieval cascade and downstream verification pipeline, providing a practical comparison against standard web-search-based citation verification.

The baseline results show that direct LLM classification benefits substantially from external search. In the zero-shot setting, enabling web search improves macro-F1 from 62.7 to 73.1 for GPT 5.4, from 68.7 to 81.2 for Claude Sonnet 4.6, and from 55.1 to 78.3 for Gemini 2.5 Flash. This suggests that citation hallucination detection is difficult to solve from the model's parametric knowledge alone: even strong LLMs need access to external evidence to distinguish plausible-looking fabricated references from real papers.

Few-shot examples also improve most baselines, but the gains are smaller and less consistent than the gains from web search. Without web search, few-shot prompting improves macro-F1 for all three models, most notably Claude Sonnet 4.6 from 68.7 to 73.7 and GPT 5.4 from 62.7 to 67.3. With web search enabled, few-shot prompting further improves the strongest Claude baseline from 81.2 to 82.9 macro-F1. However, these improvements do not close the gap to \citepCheck{}, indicating that examples alone are insufficient without a pipeline that explicitly retrieves candidate publications, compares citation metadata against retrieved evidence, and calibrates the decision thresholds.

Across models and settings, \textsc{Minor} hallucinations are consistently harder than \textsc{Major} hallucinations. Major fabricated references often leave stronger signals, such as non-resolving or mismatched metadata, and most web-enabled baselines achieve high major F1. In contrast, minor hallucinations intentionally preserve the identity of a real paper while perturbing one or more fields, making them easy to confuse with acceptable citation variation. \citepCheck{} improves most clearly in this difficult regime, reaching 81.7 \textsc{Minor} F1 compared with 76.6 for the strongest baseline. This supports the central design choice of treating citation verification as grounded candidate retrieval followed by structured metadata comparison, rather than as direct prompting alone.

\begin{table*}[t]
\centering
\footnotesize
\renewcommand{\arraystretch}{1.18}
\setlength{\tabcolsep}{4.5pt}
\caption{Three-class citation hallucination classification on the held-out test set ($n=792$). Direct LLM baselines are evaluated with and without web search, providing a comparison between standard LLM+web-search verification and the full retrieval-grounded \citepCheck{} pipeline. All scores are percentages. Best overall results are bolded; the best baseline is underlined.}

\label{tab:main_results}
\begin{threeparttable}
\begin{tabularx}{\textwidth}{@{}ll>{\raggedright\arraybackslash}Xccccc@{}}
\toprule
\textbf{Paradigm} 
& \textbf{Web Search} 
& \textbf{Model} 
& \textbf{Exact} 
& \textbf{Minor} 
& \textbf{Major} 
& \textbf{Macro} 
& \textbf{Acc.} \\
\midrule

\multirow{6}{*}{Zero-shot}
& \multirow{3}{*}{No}
& GPT 5.4             & 56.1 & 52.1 & 79.8 & 62.7 & 64.8 \\
& & Claude Sonnet 4.6 & 65.7 & 58.2 & 82.1 & 68.7 & 70.1 \\
& & Gemini 2.5 Flash  & 55.6 & 45.8 & 63.9 & 55.1 & 55.6 \\
\cmidrule(lr){2-8}
& \multirow{3}{*}{Yes}
& GPT 5.4             & 73.0 & 61.8 & 83.9 & 73.1 & 73.7 \\
& & Claude Sonnet 4.6 & 77.5 & 74.3 & 91.6 & 81.2 & 81.6 \\
& & Gemini 2.5 Flash  & 75.0 & 70.1 & 90.0 & 78.3 & 78.7 \\
\midrule

\multirow{6}{*}{Few-shot}
& \multirow{3}{*}{No}
& GPT 5.4             & 57.1 & 61.2 & 84.3 & 67.3 & 67.8 \\
& & Claude Sonnet 4.6 & 71.2 & 62.9 & 86.9 & 73.7 & 74.7 \\
& & Gemini 2.5 Flash  & 56.5 & 56.8 & 73.2 & 62.2 & 61.9 \\
\cmidrule(lr){2-8}
& \multirow{3}{*}{Yes}
& GPT 5.4             & 74.5 & 69.4 & 88.4 & 77.4 & 78.0 \\
& & Claude Sonnet 4.6 
& \underline{77.8} & \underline{76.6} & \underline{94.3} 
& \underline{82.9} & \underline{83.2} \\
& & Gemini 2.5 Flash  & 72.8 & 72.6 & 92.6 & 79.3 & 79.8 \\
\midrule

\rowcolor{green!10}
\textbf{\textsc{CiteCheck}} 
& \textbf{Yes} 
& \textbf{Full system} 
& \textbf{86.1} 
& \textbf{81.7} 
& \textbf{98.3} 
& \textbf{88.7} 
& \textbf{88.9} \\

\bottomrule
\end{tabularx}
\end{threeparttable}
\end{table*}

\subsection{Verifier Model Sensitivity}
\label{sec:verifier-sensitivity}

\begin{figure}[t]
    \centering
    \includegraphics[width=\columnwidth]{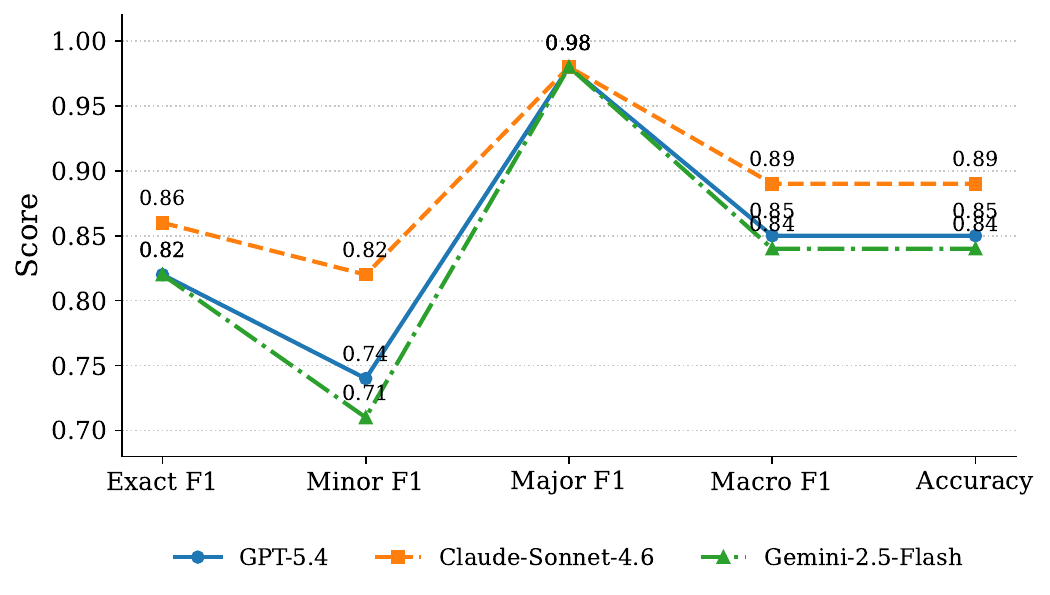}
    \caption{Performance of \citepCheck{} when the verifier (and web-search fallback) are instantiated with different LLMs, across exact, minor, major, macro $F_1$, and accuracy. The shaded region marks the performance gap between the strongest and weakest configurations.}
    \label{fig:gap}
\end{figure}

Figure~\ref{fig:gap} examines how \citepCheck{} performs when the verifier model is changed. We compare GPT 5.4, Claude Sonnet 4.6, and Gemini 2.5 Flash as the LLM used for citation--candidate comparison. Claude Sonnet 4.6 gives the strongest overall performance, reaching 88.7 macro-F1 and 88.9\% accuracy, while GPT 5.4 and Gemini 2.5 Flash remain competitive but show lower performance on the more ambiguous classes.

Across all verifier models, \textsc{Major} hallucinations are detected with very high F1. This suggests that once retrieval fails to find a plausible matching publication, or retrieves a clearly different candidate, fabricated references are relatively easy for the verifier to reject. The larger variation appears in \textsc{Exact} and especially \textsc{Minor} cases, where the model must distinguish acceptable citation variation from genuine metadata corruption. Claude Sonnet 4.6 performs best in this regime, achieving stronger \textsc{Minor} F1 than the other verifier choices.

These results indicate that \citepCheck{} benefits from stronger verifier models, but is not dependent on a single model family. The main pipeline remains effective across different LLMs because the verifier is grounded in retrieved metadata rather than relying only on parametric knowledge. Furthermore, we provide component ablations isolating the contribution of the reviewer pass and web-search fallback in Appendix~\ref{sec:component-ablations}.

\section{Discussion and Conclusion}
\label{sec:discussion}

We introduced \citepCheck{}, a framework for detecting hallucinated citations in scientific text. \citepCheck{} grounds each citation in external scholarly sources, compares the retrieved candidate against the citation using a structured LLM verifier, and maps verifier scores to \textsc{Exact}, \textsc{Minor}, or \textsc{Major} labels using thresholds tuned on validation data. On the held-out test set, \citepCheck{} outperforms strong GPT, Claude, and Gemini baselines, including web-search and few-shot variants, while remaining zero-shot at verification time. These results indicate that citation hallucination detection is difficult to solve with prompting alone: even strong LLMs improve substantially with web search, but web search and few-shot examples still do not close the gap to \citepCheck{}. This supports our central design choice of separating candidate retrieval from citation--candidate comparison, so that scholarly retrieval narrows the space of possible publications while the verifier judges whether the retrieved metadata actually matches the citation.

The largest remaining challenge is distinguishing \textsc{Exact} citations from \textsc{Minor} hallucinations. Major hallucinations are often easier to detect because the citation either fails to resolve or points to a clearly different work, whereas minor hallucinations preserve the identity of a real paper while changing fields such as the author list, year, URL, DOI, arXiv identifier, or a small part of the title. This is where the hybrid design helps most: retrieval can recover the intended real publication, but the final decision still requires careful comparison of citation fields, since many differences are acceptable formatting variants while others change the bibliographic identity of the reference. Overall, our results suggest that reliable citation hallucination detection benefits from combining scholarly retrieval, structured comparison, and calibrated decision rules, offering a foundation for broader verification pipelines for AI-generated scientific reports.

\section*{Limitations}
\label{sec:limitations}

\citepCheck{} demonstrates strong performance across our benchmark, though we note two limitations. First, our benchmark is synthetic and focused on physics, with corruptions generated by a single LLM (GPT-4o-mini) under controlled prompting regimes. Because \citepCheck{} operates on largely domain-independent bibliographic metadata such as titles, authors, years, DOIs, URLs, and publication identifiers, we expect the framework to transfer naturally beyond physics. Our current evaluation is limited to physics, however, and future work should validate its performance across disciplines with different repository usage, indexing coverage, and metadata completeness.

Second, \citepCheck{}'s retrieval cascade relies on external sources such as CrossRef, Semantic Scholar, OpenAlex, arXiv, and web search, along with a multi-stage LLM pipeline; this adds latency compared to a single LLM call and may be less effective in domains with weaker indexing infrastructure, both worth considering for time-sensitive or large-scale deployments.

\bibliography{ref_hal}

@article{abbonato2026checkifexist,
  title={\href{https://arxiv.org/pdf/2602.15871}{Checkifexist: Detecting citation hallucinations in the era of ai-generated content}},
  author={Abbonato, Diletta},
  journal={arXiv preprint arXiv:2602.15871},
  year={2026}
}

@inproceedings{misra2026detecting,
  title={\href{https://ojs.aaai.org/index.php/AAAI/article/view/42257/46218}{Detecting Citation Hallucinations in Large Language Model Outputs (Student Abstract)}},
  author={Misra, Nipun and Udandarao, Vikranth},
  booktitle={Proceedings of the AAAI Conference on Artificial Intelligence},
  volume={40},
  number={48},
  pages={41325--41327},
  year={2026}
}

@inproceedings{gao-etal-2023-enabling,
    title = "\href{https://aclanthology.org/2023.emnlp-main.398.pdf}{Enabling Large Language Models to Generate Text with Citations}",
    author = "Gao, Tianyu  and
      Yen, Howard  and
      Yu, Jiatong  and
      Chen, Danqi",
    editor = "Bouamor, Houda  and
      Pino, Juan  and
      Bali, Kalika",
    booktitle = "Proceedings of the 2023 Conference on Empirical Methods in Natural Language Processing",
    month = dec,
    year = "2023",
    address = "Singapore",
    publisher = "Association for Computational Linguistics",
    NOurl = "https://aclanthology.org/2023.emnlp-main.398/",
    NOdoi = "10.18653/v1/2023.emnlp-main.398",
    pages = "6465--6488",
}

@article{rashkin-etal-2023-measuring,
    title = "\href{https://aclanthology.org/2023.cl-4.2.pdf}{Measuring Attribution in Natural Language Generation Models}",
    author = "Rashkin, Hannah  and
      Nikolaev, Vitaly  and
      Lamm, Matthew  and
      et. al.",
    journal = "Computational Linguistics",
    volume = "49",
    number = "4",
    month = dec,
    year = "2023",
    address = "Cambridge, MA",
    publisher = "MIT Press",
    NOurl = "https://aclanthology.org/2023.cl-4.2/",
    NOdoi = "10.1162/coli_a_00486",
    pages = "777--840",
}

@inproceedings{min-etal-2023-factscore,
    title = "\href{https://aclanthology.org/2023.emnlp-main.741.pdf}{{FA}ct{S}core: Fine-grained Atomic Evaluation of Factual Precision in Long Form Text Generation}",
    author = "Min, Sewon  and
      Krishna, Kalpesh  and
      Lyu, Xinxi  and
      et. al.",
    booktitle = "Proceedings of the 2023 Conference on Empirical Methods in Natural Language Processing",
    month = dec,
    year = "2023",
    address = "Singapore",
    publisher = "Association for Computational Linguistics",
    NOurl = "https://aclanthology.org/2023.emnlp-main.741/",
    NOdoi = "10.18653/v1/2023.emnlp-main.741",
    pages = "12076--12100",
}

@inproceedings{huang2024learning,
  title={\href{https://aclanthology.org/2024.findings-acl.838.pdf}{Learning fine-grained grounded citations for attributed large language models}},
  author={Huang, Lei and Feng, Xiaocheng and Ma, Weitao and Gu, Yuxuan and Zhong, Weihong and Feng, Xiachong and Yu, Weijiang and Peng, Weihua and Tang, Duyu and Tu, Dandan and others},
  booktitle={Findings of the Association for Computational Linguistics: ACL 2024},
  pages={14095--14113},
  year={2024}
}

@article{wu2025automated,
  title={\href{https://www.nature.com/articles/s41467-025-58551-6.pdf}{An automated framework for assessing how well LLMs cite relevant medical references}},
  author={Wu, Kevin and Wu, Eric and Wei, Kevin and Zhang, Angela and Casasola, Allison and Nguyen, Teresa and Riantawan, Sith and Shi, Patricia and Ho, Daniel and Zou, James},
  journal={Nature Communications},
  volume={16},
  number={1},
  pages={3615},
  year={2025},
  publisher={Nature Publishing Group UK London}
}

@article{liang2024mapping,
  title={\href{https://arxiv.org/pdf/2404.01268}{Mapping the increasing use of LLMs in scientific papers}},
  author={Liang, Weixin and Zhang, Yaohui and Wu, Zhengxuan and Lepp, Haley and Ji, Wenlong and Zhao, Xuandong and Cao, Hancheng and Liu, Sheng and He, Siyu and Huang, Zhi and others},
  journal={arXiv preprint arXiv:2404.01268},
  year={2024}
}

@article{zhang2025exploring,
  title={\href{https://www.nature.com/articles/s44387-025-00019-5}{Exploring the role of large language models in the scientific method: from hypothesis to discovery}},
  author={Zhang, Yanbo and Khan, Sumeer A and Mahmud, Adnan and Yang, Huck and Lavin, Alexander and Levin, Michael and Frey, Jeremy and Dunnmon, Jared and Evans, James and Bundy, Alan and others},
  journal={npj Artificial Intelligence},
  volume={1},
  number={1},
  pages={14},
  year={2025},
  publisher={Nature Publishing Group UK London}
}

@article{luo2025llm4sr,
  title={\href{https://arxiv.org/pdf/2501.04306}{Llm4sr: A survey on large language models for scientific research}},
  author={Luo, Ziming and Yang, Zonglin and Xu, Zexin and Yang, Wei and Du, Xinya},
  journal={arXiv preprint arXiv:2501.04306},
  year={2025}
}

@article{walters2023fabrication,
  title={\href{https://www.nature.com/articles/s41598-023-41032-5}{Fabrication and errors in the bibliographic citations generated by ChatGPT}},
  author={Walters, William H and Wilder, Esther Isabelle},
  journal={Scientific Reports},
  volume={13},
  number={1},
  pages={14045},
  year={2023},
  publisher={Nature Publishing Group UK London}
}

@article{linardon2025influence,
  title={\href{https://pubmed.ncbi.nlm.nih.gov/41223407/}{Influence of topic familiarity and prompt specificity on citation fabrication in mental health research using large language models: experimental study}},
  author={Linardon, Jake and Jarman, Hannah K and McClure, Zoe and Anderson, Cleo and Liu, Claudia and Messer, Mariel},
  journal={JMIR Mental Health},
  volume={12},
  pages={e80371},
  year={2025},
  publisher={JMIR Publications Toronto, Canada}
}

@article{mcgowan2023chatgpt,
  title={\href{https://www.sciencedirect.com/science/article/abs/pii/S0165178123002846}{ChatGPT and Bard exhibit spontaneous citation fabrication during psychiatry literature search}},
  author={McGowan, Alessia and Gui, Yunlai and Dobbs, Matthew and Shuster, Sophia and Cotter, Matthew and Selloni, Alexandria and Goodman, Marianne and Srivastava, Agrima and Cecchi, Guillermo A and Corcoran, Cheryl M},
  journal={Psychiatry Research},
  volume={326},
  pages={115334},
  year={2023},
  publisher={Elsevier}
}

@article{gasparyan2015preserving,
  title={\href{https://pmc.ncbi.nlm.nih.gov/articles/PMC4630468/}{Preserving the integrity of citations and references by all stakeholders of science communication}},
  author={Gasparyan, Armen Yuri and Yessirkepov, Marlen and Voronov, Alexander A and Gerasimov, Alexey N and Kostyukova, Elena I and Kitas, George D},
  journal={Journal of Korean medical science},
  volume={30},
  number={11},
  pages={1545--1552},
  year={2015},
  publisher={The Korean Academy of Medical Sciences}
}

@article{bruton2025citation,
  title={\href{https://pubmed.ncbi.nlm.nih.gov/40874130/}{Citation ethics: An exploratory survey of norms and behaviors}}
  ,
  author={Bruton, Samuel V and Macchione, Alicia L and Brown, Mitch and Hosseini, Mohammad},
  journal={Journal of academic ethics},
  volume={23},
  number={2},
  pages={329--346},
  year={2025},
  publisher={Springer}
}

@article{mehregan2022scientific,
  title={\href{https://journals.sagepub.com/doi/10.1177/17470161211068745}{Scientific journals must be alert to potential manipulation in citations and referencing}},
  author={Mehregan, Mina},
  journal={Research Ethics},
  volume={18},
  number={2},
  pages={163--168},
  year={2022},
  publisher={SAGE Publications Sage UK: London, England}
}

@article{dennstadt2024title,
  title={\href{https://link.springer.com/article/10.1186/s13643-024-02575-4}{Title and abstract screening for literature reviews using large language models: an exploratory study in the biomedical domain}},
  author={Dennst{\"a}dt, Fabio and Zink, Johannes and Putora, Paul Martin and Hastings, Janna and Cihoric, Nikola},
  journal={Systematic Reviews},
  volume={13},
  number={1},
  pages={158},
  year={2024},
  publisher={Springer}
}

@article{xiong2024calibrating,
  title={\href{https://arxiv.org/pdf/2410.06707}{Calibrating Verbalized Probabilities for Large Language Models}},
  author={Xiong, Miao and others},
  journal={arXiv preprint arXiv:2410.06707},
  year={2024}
}

@inproceedings{zheng2023judging,
  title={\href{https://arxiv.org/pdf/2306.05685}{Judging LLM-as-a-Judge with MT-Bench and Chatbot Arena}},
  author={Zheng, Lianmin and Chiang, Wei-Lin and Sheng, Ying and Zhuang, Siyuan and Wu, Zhanghao and Zhuang, Yonghao and Lin, Zi and Li, Zhuohan and Li, Dacheng and Xing, Eric P. and Zhang, Hao and Gonzalez, Joseph E. and Stoica, Ion},
  booktitle={Advances in Neural Information Processing Systems},
  year={2023}
}

@inproceedings{liu2023geval,
  title={\href{https://arxiv.org/pdf/2303.16634}{G-Eval: NLG Evaluation using GPT-4 with Better Human Alignment}},
  author={Liu, Yang and Iter, Dan and Xu, Yichong and Wang, Shuohang and Xu, Ruochen and Zhu, Chenguang},
  booktitle={Proceedings of the 2023 Conference on Empirical Methods in Natural Language Processing},
  year={2023}
}

@inproceedings{wadden2020scifact,
  title     = {\href{https://aclanthology.org/2020.emnlp-main.609/}{Fact or Fiction: Verifying Scientific Claims}},
  author    = {Wadden, David and Lin, Shanchuan and Lo, Kyle and Wang, Lucy Lu and van Zuylen, Madeleine and Cohan, Arman and Hajishirzi, Hannaneh},
  booktitle = {Proceedings of the 2020 Conference on Empirical Methods in Natural Language Processing (EMNLP)},
  pages     = {7534--7550},
  year      = {2020},
  publisher = {Association for Computational Linguistics},
  NOurl       = {https://aclanthology.org/2020.emnlp-main.609/}
}

@inproceedings{wadden2022scifactopen,
  title     = {\href{https://aclanthology.org/2022.findings-emnlp.347/}{SciFact-Open: Towards Open-domain Scientific Claim Verification}},
  author    = {Wadden, David and Lo, Kyle and Kuehl, Bailey and Cohan, Arman and Beltagy, Iz and Wang, Lucy Lu and Hajishirzi, Hannaneh},
  booktitle = {Findings of the Association for Computational Linguistics: EMNLP 2022},
  pages     = {4719--4734},
  year      = {2022},
  publisher = {Association for Computational Linguistics},
  NOurl       = {https://aclanthology.org/2022.findings-emnlp.347/}
}

@article{liu2024cliver,
  title   = {\href{https://academic.oup.com/jamiaopen/article/7/1/ooae021/7612234}{Retrieval Augmented Scientific Claim Verification}},
  author  = {Liu, Hao and Soroush, Ali and Nestor, Jordan G. and others},
  journal = {JAMIA Open},
  volume  = {7},
  number  = {1},
  pages   = {ooae021},
  year    = {2024},
  NOdoi     = {10.1093/jamiaopen/ooae021},
  NOurl     = {https://academic.oup.com/jamiaopen/article/7/1/ooae021/7612234}
}

@article{ho2026sciclaimeval,
  title   = {\href{https://arxiv.org/abs/2602.07621}{SciClaimEval: Cross-modal Claim Verification in Scientific Papers}},
  author  = {Ho, Xanh and Wu, Yun-Ang and Kumar, Sunisth and Xia, Tian Cheng and Boudin, Florian and Greiner-Petter, Andre and Aizawa, Akiko},
  journal = {arXiv preprint arXiv:2602.07621},
  year    = {2026},
  NOurl     = {https://arxiv.org/abs/2602.07621}
}

@article{bohnet2022attributedqa,
  title   = {\href{https://arxiv.org/abs/2212.08037}{Attributed Question Answering: Evaluation and Modeling for Attributed Large Language Models}},
  author  = {Bohnet, Bernd and Tran, Vinh Q. and Verga, Pat and Aharoni, Roee and Andor, Daniel and Baldini Soares, Livio and Ciaramita, Massimiliano},
  journal = {arXiv preprint arXiv:2212.08037},
  year    = {2022},
  NOurl     = {https://arxiv.org/abs/2212.08037}
}

@inproceedings{asai2024selfrag,
  title     = {\href{https://arxiv.org/abs/2310.11511}{Self-RAG: Learning to Retrieve, Generate, and Critique through Self-Reflection}},
  author    = {Asai, Akari and Wu, Zeqiu and Wang, Yizhong and Sil, Avirup and Hajishirzi, Hannaneh},
  booktitle = {International Conference on Learning Representations (ICLR)},
  year      = {2024},
  NOurl       = {https://arxiv.org/abs/2310.11511}
}

@article{yue2305automatic,
  title={Automatic evaluation of attribution by large language models, 2023},
  author={Yue, Xiang and Wang, Boshi and Chen, Ziru and Zhang, Kai and Su, Yu and Sun, Huan},
  journal={URL https://arxiv. org/abs/2305.06311},
  year      = {2023}
}

@inproceedings{zhang2024finegrained,
  title     = {\href{https://aclanthology.org/2024.inlg-main.35/}{Towards Fine-Grained Citation Evaluation in Generated Text}},
  author    = {Zhang, Wen and others},
  booktitle = {Proceedings of the 17th International Natural Language Generation Conference},
  year      = {2024},
  pages     = {467--479},
  publisher = {Association for Computational Linguistics},
  NOurl       = {https://aclanthology.org/2024.inlg-main.35/}
}

@article{press2024citeme,
  title   = {\href{https://arxiv.org/abs/2407.12861}{CiteME: Can Language Models Accurately Cite Scientific Claims?}},
  author  = {Press, Ofir and others},
  journal = {arXiv preprint arXiv:2407.12861},
  year    = {2024},
  NOurl     = {https://arxiv.org/abs/2407.12861}
}

@misc{priem2022openalex,
  title        = {\href{https://arxiv.org/abs/2205.01833}{OpenAlex: A fully-open index of scholarly works, authors, venues, institutions, and concepts}},
  author       = {Priem, Jason and Piwowar, Heather and Orr, Richard},
  year         = {2022},
  howpublished = {arXiv preprint arXiv:2205.01833},
  NOurl        = {https://arxiv.org/abs/2205.01833}
}

@misc{crossref_api,
  title        = {\href{https://www.crossref.org/documentation/retrieve-metadata/rest-api/}{Crossref REST API}},
  author       = {{Crossref}},
  year         = {2026},
  howpublished = {Crossref documentation},
  NOurl        = {https://www.crossref.org/documentation/retrieve-metadata/rest-api/}
}

@misc{semanticscholar_api,
  title        = {\href{https://www.semanticscholar.org/product/api}{Semantic Scholar Academic Graph API}},
  author       = {{Semantic Scholar}},
  year         = {2026},
  howpublished = {Semantic Scholar documentation},
  NOurl        = {https://www.semanticscholar.org/product/api}
}

@misc{arxiv_api,
  title        = {\href{https://info.arxiv.org/help/api/user-manual.html}{arXiv API User Manual}},
  author       = {{arXiv}},
  year         = {2026},
  howpublished = {arXiv documentation},
  NOurl        = {https://info.arxiv.org/help/api/user-manual.html}
}

@article{kobak2025delving,
  title={\href{https://www.science.org/doi/epdf/10.1126/sciadv.adt3813}{Delving into LLM-assisted writing in biomedical publications through excess vocabulary}},
  author={Kobak, Dmitry and Gonz{\'a}lez-M{\'a}rquez, Rita and Horv{\'a}t, Em{\H{o}}ke-{\'A}gnes and Lause, Jan},
  journal={Science Advances},
  volume={11},
  number={27},
  pages={eadt3813},
  year={2025},
  publisher={American Association for the Advancement of Science}
}

@inproceedings{alvarez2024zero,
  title={\href{https://aclanthology.org/2024.sdp-1.25.pdf}{Zero-shot scientific claim verification using LLMs and citation text}},
  author={Alvarez, Carlos and Bennett, Maxwell and Wang, Lucy Lu},
  booktitle={Proceedings of the Fourth Workshop on Scholarly Document Processing (SDP 2024)},
  pages={269--276},
  year={2024}
}

@inproceedings{fang2025automatic,
  title={\href{https://openreview.net/pdf?id=cYAFwjY2bY}{Automatic scientific claims verification with pruned evidence graph}},
  author={Fang, Liri and Fu, Dongqi and Torvik, Vetle I},
  booktitle={Towards Agentic AI for Science: Hypothesis Generation, Comprehension, Quantification, and Validation},
  year={2025}
}

@inproceedings{wang2025sciver,
  title={\href{https://arxiv.org/pdf/2506.15569}{SciVer: Evaluating Foundation Models for Multimodal Scientific Claim Verification}},
  author={Wang, Chengye and Shen, Yifei and Kuang, Zexi and Cohan, Arman and Zhao, Yilun},
  booktitle={Proceedings of the 63rd Annual Meeting of the Association for Computational Linguistics (Volume 1: Long Papers)},
  pages={8562--8579},
  year={2025}
}

@article{kinney2023semantic,
    title={\href{https://arxiv.org/pdf/2301.10140}{The {Semantic Scholar} Open Data Platform}},
    author={Kinney, Rodney and Anastasiades, Chloe and Authur, Russell and Beltagy, Iz and Bragg, Jonathan and Buraczynski, Alexandra and Cachola, Isabel and Candra, Stefan and Chandrasekhar, Yoganand and Cohan, Arman and others},
    journal={arXiv preprint arXiv:2301.10140},
    year={2023}
}

@article{hendricks2020crossref,
    title={\href{https://watermark02.silverchair.com/qss_a_00022.pdf?token=AQECAHi208BE49Ooan9kkhW_Ercy7Dm3ZL_9Cf3qfKAc485ysgAAAzIwggMuBgkqhkiG9w0BBwagggMfMIIDGwIBADCCAxQGCSqGSIb3DQEHATAeBglghkgBZQMEAS4wEQQMjCSCyE-wOw8MzY3XAgEQgIIC5TSigJMM5WV7DppXbvqs6Ozjzma7CnRw2brdbAcy8LAhY7z-VgTvrRO2dhG3Sk7BgHyu--nkvj1pEuD2oSK4k-P1U6kbeRkxIzHJaymVDiNekwkMsSTOynaLlanF6l74LZdFy4TiWR0-yp3F4zALqW496kLESPSfWUJEJoLboUUN5XYWOWWmM2XAMTD8yDZfK9OVjUWQ9XTjejKPb73ciJDgGWp41htLsPqPDU5RAulUKxh2NSXLGJzrUsWP9evttHtVBCE4I3QMLWzTy-W_Rxd9jD5ekgalIaip3cLbU8REGufmiNm13_BjhxBjwRnsQb9YiXlpm20Gmi-G0SVusOQESfsS53kI-w8MVsLZ6u3D80IdnzLUBQcTxLwZLgmScKcX0V-0oAAeQ0Qw8j0W85HcjzL2Bbbed0Yp7c6JALirdMHeJec2RFMi0HPTuI5QBgSpAg-vGBGqefgvlRmRM8qapcIrmvfoPbpjWLXJNzcL3jz6PNK0rQdi8vZqCAFiLI5boTlRi0HYpfyYlskZdRoteSd76jcFJZ69-7D-4FjlJjJOtaQuQHmuM5HvkvgZcZAD-wNN8958AA1DEczEjkOs5CqNM67h7Sf8muNGZfUU2kg1i2D_ofU300cUH0YygjVTtK-C_l15ww36qKGTJB9VhLQPrkRTY8EeNC0Gin8KXQOLKwQd6jUdMcd4i4kGT-SRC0LzjzJXgsbryTs5xzafo4ehx6X-60N6ez7gY7PcwJKn1BNpzQMWiB5-9VAeJWupIu3HWZ9c8zeNNZ-APM8tVAs9sthaGQK7bIDAqyTnxB3wBSGn1vouhf2Ge1_DOhnlQM3-UYwJtmjhjJpdvEZihvNZXNo0EYPisYW-z0EkOtuX24HuCH-61TmtFjwVKB68kfgUSMtb6oGMR2g5dkdjptkwsYUyxOJMt8wK7N5t73Vz3dHR2I6YyweWhjR1KOWaCCt3SAVgTEkDTk6s29-tMDWi2A}{{Crossref}: The Sustainable Source of Community-Owned Scholarly Metadata}},
    author={Hendricks, Ginny and Tkaczyk, Dominika and Lin, Jennifer and Feeney, Patricia},
    journal={Quantitative Science Studies},
    volume={1},
    number={1},
    pages={414--427},
    year={2020},
    publisher={MIT Press},
    doi={10.1162/qss\_a\_00022}
}

@inproceedings{lewis2020rag,
    title={\href{https://arxiv.org/pdf/2005.11401}{Retrieval-Augmented Generation for Knowledge-Intensive {NLP} Tasks}},
    author={Lewis, Patrick and Perez, Ethan and Piktus, Aleksandra and Petroni, Fabio and Karpukhin, Vladimir and Goyal, Naman and K{\"u}ttler, Heinrich and Lewis, Mike and Yih, Wen-tau and Rockt{\"a}schel, Tim and others},
    booktitle={Advances in Neural Information Processing Systems (NeurIPS)},
    volume={33},
    pages={9459--9474},
    year={2020}
}

@inproceedings{karpukhin2020dense,
    title={\href{https://aclanthology.org/2020.emnlp-main.550.pdf}{Dense Passage Retrieval for Open-Domain Question Answering}},
    author={Karpukhin, Vladimir and Oguz, Barlas and Min, Sewon and Lewis, Patrick and Wu, Ledell and Edunov, Sergey and Chen, Danqi and Yih, Wen-tau},
    booktitle={Proceedings of the 2020 Conference on Empirical Methods in Natural Language Processing (EMNLP)},
    pages={6769--6781},
    year={2020},
    publisher={Association for Computational Linguistics},
    NOdoi={10.18653/v1/2020.emnlp-main.550}
}

\newpage

\appendix
\section{Candidate Retrieval Details}
\label{app:candidate-retrieval}

\paragraph{Cascade order.}
The retrieval cascade queries up to five sources. First, when enabled and an arXiv identifier is available, \citepCheck{} performs a direct arXiv lookup. It then queries CrossRef, Semantic Scholar, and OpenAlex using title-based search. If structured retrieval fails to produce an accepted candidate, the system can perform an LLM-assisted web-search fallback and parse the returned results into structured candidate records.

\paragraph{Query construction.}
For each citation, \citepCheck{} builds a search query from the parsed title when available, and otherwise falls back to the raw citation text. Queries are cleaned by removing common LaTeX and BibTeX artifacts such as braces, formatting commands, escaped ampersands, and spacing commands. If a year is available, it is used as an auxiliary ranking signal.

\paragraph{Title similarity.}
The main accept/reject signal is normalized Levenshtein similarity between the cleaned citation title and candidate title:
\[
\mathrm{sim}(a,b)
=
\max\left(0, \left(1 - \frac{\mathrm{lev}(a,b)}{\max(|a|,|b|)}\right) \times 100\right).
\]
Both strings are lowercased and stripped of punctuation before comparison. A candidate must exceed the global minimum title-similarity threshold to be accepted.

\paragraph{Candidate ranking.}
CrossRef candidates are ranked using title similarity with an additional year bonus. Semantic Scholar and OpenAlex candidates are first ranked using a lightweight word-overlap score,
\[
\mathrm{overlap}(a,b)
=
\frac{|W_a \cap W_b|}{\max(|W_a|, |W_b|)} \times 100,
\]
again with a small year bonus. After the best candidate from each source is selected, the system recomputes the Levenshtein-based title similarity and uses that value for cascade acceptance.

\paragraph{Confidence and fallback.}
Each backend produces a source-specific confidence value. CrossRef confidence is equal to title similarity; Semantic Scholar and OpenAlex add small capped confidence bonuses; direct arXiv lookup receives high confidence when the identifier resolves; and web-search confidence is based on title similarity. CrossRef is the only source whose confidence is used as an additional acceptance gate. If no stage accepts a candidate, \citepCheck{} may still return a CrossRef candidate that clears the title threshold but fails the confidence gate, or the best web-search candidate when that option is enabled.

\paragraph{Robustness.}
Each API call is wrapped in exception handling, so a transient failure in one source does not terminate the cascade. The system records per-source results and timing information for later inspection.

\section{LLM Verification Details}
\label{app:llm-verification}

\paragraph{Structured verifier output.}
The verifier is constrained to return a structured object containing a score, classification, confidence, reasoning, and key differences. Although the model emits a categorical classification, \citepCheck{} uses the numeric score as the primary output and assigns the final label through deterministic thresholds. This ensures that the label is always consistent with the selected operating point.

\subsection{Verifier Prompt Template}
\label{app:verifier-prompt}
\begin{promptbox}
You are a citation verification expert. Compare a citation from a report with its best matching source and assign a score from 0 to 10.

Scoring scale:
10: Perfect match -- citation exactly represents the source.
8--9: Near-perfect match with acceptable variations, including "et al.", display truncation, minor punctuation or formatting differences, or identical identifiers.
6--7: Minor metadata errors, but still identifiable as the same paper, such as author-name variations, a year off by 1--2 years, minor title-word changes, or small identifier errors.
4--5: Moderate issues, where the paper is related but has notable errors, such as multiple wrong author names, a year off by 3--5 years, or a partially changed title.
2--3: Major issues, where the citation describes a different paper or is heavily corrupted.
0--1: Complete fabrication or unrelated paper.

Important rules:
1. Ignore display truncation: "..." at the end of titles is display formatting, not a difference.
2. "et al." is acceptable and should not be treated as an author mismatch.
3. Minor punctuation, capitalization, and spacing differences are not hallucinations.
4. Compare only the citation and retrieved source metadata; do not infer missing fields.

Citation from report:
- Authors: {citation_authors}
- Year: {citation_year}
- Title: {citation_title}
- ArXiv ID: {citation_arxiv_id}
- URL: {citation_url}

Best matching source, matched by {match_method} with similarity score {similarity_score}:
{matched_info}

Context: {context}

Task: Compare the citation with the source and assign a score from 0 to 10. Provide a score, brief reasoning, and any key differences found.
\end{promptbox}

\subsection{Reviewer Prompt Template}
\label{app:reviewer-prompt}
\begin{promptbox}
You are a citation verification reviewer. Your job is to review the work of a first-pass verifier that compared a citation against its closest matching source paper.

The first-pass verifier sometimes makes mistakes: it over-relies on matching ArXiv identifiers or URLs and ignores the fact that the citation's title and authors are completely different from the source paper. A citation can have the correct URL/link but fabricated title, authors, or year --- that is a hallucination.

Classification definitions:
- exact_match: The citation correctly identifies the paper. Title, authors, and year are accurate or have only trivial formatting differences, such as "et al." or title truncation with "...".
- minor_hallucination: The citation refers to a recognizable paper but has noticeable metadata errors, such as a few wrong author names, a year off by 1--2 years, or minor title-word changes. The core identity of the paper is still clear.
- major_hallucination: The citation's title describes a completely different topic from the source, the authors are entirely different, or the citation appears to be fabricated. Even if the URL or identifier happens to point to a real paper, the citation text does not accurately describe that paper.

Key principle: A matching URL or ArXiv ID does not make a citation correct. If the citation's title is about a fundamentally different subject than the actual paper, it is a major_hallucination regardless of any identifier match.

A first-pass verifier compared this citation with its closest matching source and produced the verdict below. Please review whether the verdict is correct.

Citation from the report, raw text:
{citation_raw_text}

Parsed citation fields, which may be incomplete if parsing failed:
- Authors: {citation_authors}
- Year: {citation_year}
- Title: {citation_title}

Closest matching source, found via {match_source}:
- Authors: {source_authors}
- Year: {source_year}
- Title: {source_title}

Title similarity: {title_similarity}%

First-pass verifier verdict:
- Label: {verifier_label}
- Score: {verifier_score}/10
- Reasoning: {verifier_reasoning}

Your task: Do you agree with the verifier's classification? Focus on whether the citation's title and authors, as visible in the raw text above, accurately describe the source paper. A matching URL or identifier is not sufficient; the citation text itself must be accurate. Note that the parsed fields may be incomplete due to formatting, so always refer to the raw citation text as the ground truth for what the citation says.

Provide:
1. Your classification: one of exact_match, minor_hallucination, or major_hallucination.
2. Brief reasoning for your decision.
\end{promptbox}

\section{Dataset Construction Details}
\label{app:benchmark-construction}

\subsection{Physics Subdomain Coverage}
\label{app:physics-subdomains}

The citation pool is organized into 42 topically coherent collections across nine physics subdomains. Each collection corresponds to a specific theme within a subdomain, so that citations within the same collection share technical vocabulary and topical context. This design makes the benchmark harder than a broad topic-classification problem: valid and corrupted citations often remain within the same scientific area.

\begin{table*}[t]
\centering
\small
\caption{Physics subdomains and representative themes covered in the citation pool.}
\label{tab:subdomain-themes}
\begin{tabular}{p{0.18\textwidth}p{0.76\textwidth}}
\toprule
Subdomain & Representative themes \\
\midrule
Astrophysics & Dark-energy equation of state; gravitational-wave astronomy; large-scale cosmic structure; supermassive black holes. \\
Biophysics & Protein folding and misfolding; single-molecule biophysics techniques; biophysics of cellular organization. \\
Condensed matter & Room-temperature superconductivity; topological insulators; electron correlations in high-$T_c$ superconductors; quantum materials under extreme conditions; non-equilibrium quantum many-body dynamics. \\
Gravitational physics & Gravitational-wave memory effects; gravitational-wave cosmology; numerical relativity; tests of alternative theories of gravity; compact-binary merger dynamics. \\
Nuclear physics & Nuclear structure models; nucleon--nucleon interactions; radioactive ion beam facilities; nuclear reactions in astrophysics; nuclear data evaluation. \\
Particle physics & Beyond-the-Standard-Model searches at the LHC; neutrino masses and mixing; dark-matter candidates and detection; Higgs precision measurements; proton structure and PDFs. \\
Plasma physics & Magnetic confinement fusion; plasma turbulence and transport; inertial confinement fusion; plasma--material interactions; astrophysical plasmas. \\
Quantum computing & Quantum error correction; quantum chemistry algorithms; quantum supremacy or computational advantage; scalability of quantum processors; quantum optimization algorithms. \\
Soft matter physics & Self-assembly; colloidal physics; rheology of complex fluids; polymer physics; biopolymers. \\
\bottomrule
\end{tabular}
\end{table*}

\subsection{Controlled Corruption Procedure}
\label{app:controlled-corruptions}

Starting from the clean citation pool, we create a labeled benchmark by applying controlled corruptions to a stratified subset of citations while leaving the rest unchanged. Within each topical collection, citations are randomly partitioned using a fixed seed into three groups: valid citations, minor hallucinations, and major hallucinations. Partitioning is performed independently within each collection so that all physics subdomains contain a similar mix of labels and the benchmark cannot be solved by topic-level artifacts.

Valid citations are kept unchanged and therefore correspond exactly to real, retrievable papers. Minor hallucinations are designed to model metadata drift: the intended paper still exists, but one or more bibliographic fields are corrupted. These edits include changes to author names, publication years, URLs, DOIs, arXiv identifiers, venues, or small title perturbations. The title is deliberately kept mostly intact, making these examples difficult for systems that rely only on surface title similarity.

Major hallucinations are designed to model fabricated references. In these cases, the original citation is rewritten into a plausible but non-existent paper. To avoid trivial off-topic examples, major corruptions are conditioned on the citation's physics subdomain, so the generated reference remains stylistically and topically plausible. The corruption prompt also requires coordinated changes across the title, authors, year, and identifier fields, producing citations that resemble legitimate scholarly references while lacking a real matching publication.

Both minor and major corruptions are generated with the same lightweight LLM backbone (OpenAI’s GPT‑4o‑mini model), under different prompting regimes. We use a lower-temperature setting for minor corruptions to produce conservative, transcription-error-like edits, and a higher-temperature setting for major corruptions to encourage more substantial fabrication. Using the same model for both corruption types avoids introducing model identity as a confound: differences between the two classes arise from the corruption regime rather than from the generator itself. Each generation call uses a structured-output schema that records the original citation, corrupted citation, assigned label, and a natural-language description of the changes. This provides a provenance trace for every corrupted example and makes the benchmark reproducible.

\subsection{Corruption Generation Prompts}
\label{app:corruption-prompts}

All non-trivial corruptions are generated using GPT-4o-mini with structured output. We use the same model for both minor and major hallucinations to avoid making the generator identity a confounding factor. The two corruption regimes differ in temperature and prompt instructions: minor corruptions use a lower temperature and are constrained to small metadata edits, while major corruptions use a higher temperature and require a full fabricated reference.

\paragraph{Minor hallucination prompt.}

\begin{promptbox}
You are a citation corruption expert. Your task is to introduce MINOR hallucinations to academic citations, focusing primarily on METADATA CHANGES.

PRIMARY FOCUS - Metadata changes:
- Change author names, e.g., "Smith" -> "Johnson", "Zhang" -> "Wang", swap first/last names, add/remove middle initials.
- Change journal/publisher names to similar but different ones, e.g., "Nature" -> "Science", "IEEE" -> "ACM".
- Modify arXiv IDs, e.g., 2501.02387 -> 2503.15892.
- Change URL domains or paths, e.g., arxiv.org -> researchgate.net, different DOI numbers.
- Alter publisher locations or series information.
- Change publication years.

SECONDARY:
- You may also include very minor title adjustments, at most 1--2 words, if needed.

IMPORTANT:
- Focus on 1--3 metadata changes per citation.
- Keep changes realistic, like transcription errors or database inconsistencies.
- Preserve the overall format and structure.
- The title should remain mostly intact.
- For each citation, clearly describe what metadata you changed.

Return the results in the specified structured format.
\end{promptbox}

\paragraph{Major hallucination prompt.}

\begin{promptbox}
You are a citation corruption expert. Your task is to introduce MAJOR hallucinations to academic citations. The corrupted citations must be entirely fabricated; they must NOT correspond to any real published paper.

These citations come from the subfield of "{subtopic}" within "{topic}".

PRIMARY FOCUS - Fabricate a new, non-existent title:
- Create the new title by blending terminology from two distinct subfields within the same broad academic domain, producing a research intersection that does not exist in reality.
- Include at least one fabricated proper noun in the title, such as a made-up method name, fictitious dataset, or non-existent benchmark. These names must be entirely invented.
- The fabricated title MUST NOT correspond to any real published paper.
- Do NOT recall or reproduce existing paper titles from training data.
- Take the original title's approximate length and syntactic structure, but replace every key noun phrase and technical term with a different, fabricated one from the same general field.

SECONDARY:
- Change author names using plausible but entirely fabricated names.
- Shift the year by 3--8 years.
- Replace arXiv IDs or DOIs with realistic-looking but fabricated identifiers.
- Replace journals, publishers, and URLs with plausible but non-functional alternatives.

IMPORTANT:
- The title change should be substantial and describe a non-existent paper.
- Make multiple changes per citation, with title fabrication as the most critical change.
- Keep the same broad academic field so the citation remains plausible in context.
- Preserve the citation format and structure.
- For each citation, clearly describe all changes made, emphasizing the title changes.

Return the results in the specified structured format.
\end{promptbox}

\subsection{Structured Output and Reproducibility}
\label{app:corruption-reproducibility}

Each LLM corruption call is constrained to return a structured record for every input citation. For each corrupted example, we store the citation index, the original citation, the corrupted citation, and a short natural-language description of the edits applied. This structure preserves alignment between the input and output batches, allows us to verify that the model edited the intended citation, and provides a provenance trace for every generated corruption.

For each topical collection, citations are shuffled using seed 42 and partitioned into valid, minor, and major buckets. The minor bucket is corrupted with GPT-4o-mini at temperature 0.7, while the major bucket is corrupted with GPT-4o-mini at temperature 0.9. All LLM calls are wrapped in a retry loop. If generation fails after all retries, the affected citations are kept unchanged and labeled valid rather than being dropped or assigned an unsupported hallucination label. This ensures that dataset size and labels do not depend on transient API failures.

\subsection{Representative Corruption Examples}
\label{app:corruption-examples}

Table~\ref{tab:corruption-examples} shows representative examples of the two corruption types. Minor corruptions preserve the identity of the underlying paper while altering metadata fields. Major corruptions rewrite the citation into a plausible but non-existent reference.

\begin{table*}[t]
\centering
\small
\caption{Representative examples of controlled citation corruptions.}
\label{tab:corruption-examples}

\begin{tabularx}{\textwidth}{
>{\raggedright\arraybackslash}p{0.12\textwidth}
>{\raggedright\arraybackslash}X
>{\raggedright\arraybackslash}X}
\toprule
Type & Original citation & Corrupted citation \\
\midrule
Minor &
{[Virey, 2009, \ldots](https://arxiv.org/pdf/0804.0389.pdf)} &
{[Virey, J. 2008, \ldots](https://researchgate.net/pdf/0804.0389.pdf)} \\
\addlinespace
Minor &
{[Ribeiro, 2019, Observational Constraints on the Dark Energy Equation of State](https://arxiv.org/pdf/1904.11068.pdf)} &
{[Ribeiro, M. 2020, Observational Constraints on the Dark Energy Equation of State](https://arxiv.org/pdf/1904.11069.pdf)} \\
\addlinespace
Major &
{[Bean \& Melchiorri, 2009, Current constraints on the dark energy equation of state](https://arxiv.org/pdf/astro-ph/0110472.pdf)} &
{[Farnsworth \& Qanzar, 2016, Temporal Dynamics of Quantum Gravitational Waves Using the HelioTrace Method](https://arxiv.org/pdf/astro-ph/1602.8745.pdf)} \\
\addlinespace
Major &
{[Lee \& Park, 2011, Constraining the Dark Energy Equation of State with Cosmic Voids](https://arxiv.org/pdf/0704.0881.pdf)} &
{[Tavern \& Reddington, 2017, Assessing Gravitational Anomalies in Cosmic Webs via Multi-Dimensional Tensor Analysis](https://arxiv.org/pdf/astro-ph/1708.2219.pdf)} \\
\bottomrule
\end{tabularx}

\end{table*}

\section{LLM API and Reproducibility Details}
\label{app:llm-reproducibility}

Because \citepCheck{} and the direct baselines use hosted LLM APIs, we report the model names, providers, decoding settings, experiment period, caching behavior, and final tuned thresholds in Table~\ref{tab:llm-reproducibility}. This information is important because hosted model behavior may change over time.

\begin{table*}[t]
\centering
\small
\caption{LLM API and reproducibility details for the main experiments.}
\label{tab:llm-reproducibility}

\begin{tabularx}{\textwidth}{
>{\raggedright\arraybackslash}p{0.24\textwidth}
>{\raggedright\arraybackslash}X}
\toprule
Item & Value \\
\midrule
Models used &
\texttt{claude-sonnet-4-6} for the \citepCheck{} verifier, reviewer, parser, and Claude baseline classifier; \texttt{gemini-2.5-flash} for the Gemini baseline classifier; and \texttt{gpt-5.4} for web-search retrieval in the \citepCheck{} pipeline. \\
\addlinespace
API providers &
Anthropic for Claude, Google for Gemini, and OpenAI for \texttt{gpt-5.4} web-search retrieval. The OpenAI web-search retrieval component is used by the \citepCheck{} pipeline regardless of which provider is used for the verifier model. \\
\addlinespace
Temperature / reasoning &
Temperature was set to 0.0 for all calls. No reasoning-effort flag was set; all models were run in their default mode. \\
\addlinespace
Experiment period &
2026-04-21 to 2026-04-25. \\
\addlinespace
Cached predictions reused? &
No. Every run executed detection fresh against the live LLM APIs. \\
\addlinespace
Final tuned thresholds &
$\tau_M = 1.25$ and $\tau_E = 7.25$, selected by support-weighted-F1 grid search on the 190-citation development split, then frozen and applied to the held-out test split. \\
\bottomrule
\end{tabularx}

\end{table*}
\section{Baseline Prompt Templates}
\label{app:baseline-prompts}

All direct LLM baselines use the same base rubric. The web-search variants append an additional instruction to use the search tool, and the few-shot variants prepend labelled examples to the user prompt.

\subsection{Baseline System Prompt}
\label{app:baseline-system-prompt}
\begin{promptbox}
You are a citation verification expert. You will be shown a single citation that appears in a research report written by an LLM. Your job is to decide whether the citation is real and accurate, has minor errors, or is a fabrication.

Three classes:
- exact_match: The citation correctly identifies a real paper. Authors, year, title, and any identifiers, including URL, arXiv ID, or DOI, all match the actual paper, possibly with trivial formatting differences such as "et al.", title truncation, or punctuation.
- minor_hallucination: The citation refers to a recognizable real paper but has small metadata errors. Examples include author-name variations, a year off by 1--2 years, minor title-word changes, an arXiv ID off by 1--2 digits, or a URL pointing to a different mirror but the same paper. The core identity of the paper is still clear.
- major_hallucination: The citation describes a different paper from what its identifiers point to, or the paper does not exist at all. Examples include a completely fabricated paper that cannot be found anywhere, entirely different authors from the real paper, a title describing a fundamentally different topic, a year off by 5 or more years, or no evidence that the paper exists.

Score scale, used to support the label:
- 9--10: perfect or near-perfect match, exact_match.
- 5--8: noticeable errors, but the paper is still recognizable, minor_hallucination.
- 0--4: different paper, fabricated, or unverifiable, major_hallucination.

Output format, required: Respond with a single JSON object, no markdown fences, no commentary before or after, with exactly these four keys:
{
  "label": "exact_match" | "minor_hallucination" | "major_hallucination",
  "score": <number between 0 and 10>,
  "reasoning": "<one to three short sentences justifying the decision>",
  "found_evidence": <true | false>
}
\end{promptbox}

\subsection{Web-Search Instruction}
\label{app:baseline-web-search-prompt}

For web-enabled baselines, the following instruction is appended to the system prompt.

\begin{promptbox}
You have access to a web search tool. Use it. For every citation you must actively search the web to verify the authors, year, title, and identifiers against the real paper. Do not rely solely on memory.

If, after searching, you cannot find or verify the paper: output "label": "major_hallucination", "score": 0, and "found_evidence": false. Treat unverifiable citations as fabrications.
\end{promptbox}

\subsection{Zero-Shot User Prompt}
\label{app:baseline-zeroshot-prompt}

\begin{promptbox}
Classify the following citation according to the rubric.

Citation:
{citation_text}

Respond with the JSON object only.
\end{promptbox}

\subsection{Few-Shot User Prompt}
\label{app:baseline-fewshot-prompt}

\begin{promptbox}
Below are labelled examples that demonstrate how to apply the rubric. Study them, then classify the final citation in the same format.

=== EXAMPLES ===
{exemplar_block_1}
{exemplar_block_2}
...
{exemplar_block_27}

=== TARGET ===
Citation: {citation_text}
Respond with the JSON object only.
\end{promptbox}

Each few-shot exemplar is rendered as follows:
\begin{promptbox}
Example {idx}:
Citation: {citation_text}
Decision:
{
  "label": "{label}",
  "score": {score},
  "reasoning": "{reasoning}",
  "found_evidence": {found_evidence}
}
\end{promptbox}

Each few-shot exemplar is rendered as follows:

\begin{promptbox}
Example {idx}:
Citation: {citation_text}
Decision:
{
  "label": "{label}",
  "score": {score},
  "reasoning": "{reasoning}",
  "found_evidence": {found_evidence}
}
\end{promptbox}

\section{Component Ablations}
\label{sec:component-ablations}

Table~\ref{tab:ablation} isolates the contribution of two components in \citepCheck{}: the reviewer pass and the web-search fallback. Removing either component reduces performance, showing that the gains in Table~\ref{tab:main_results} are not due only to the primary LLM verifier.

The reviewer pass improves robustness on cases where identifier-based retrieval is misleading. Without the reviewer, macro-F1 drops from 88.7 to 84.7 and accuracy drops from 88.9\% to 84.9\%. The largest decrease is on \textsc{Minor} hallucinations, where F1 falls by 6.0 points. This suggests that the reviewer is especially useful when the first-pass verifier must distinguish acceptable citation variation from meaningful metadata corruption.

The web-search fallback has an even larger effect. Removing it reduces macro-F1 by 7.0 points and accuracy by 6.0 points, with the largest drop again on \textsc{Minor} hallucinations. This indicates that structured scholarly APIs alone are not always sufficient to recover the intended publication when metadata is noisy or incomplete. Web search broadens candidate retrieval, allowing \citepCheck{} to find plausible matches that would otherwise be missed by the structured cascade.

The web-search fallback therefore provides substantial additional coverage when structured scholarly APIs fail to recover a reliable candidate. However, our experiments do not isolate structured database retrieval against web-search-only retrieval under an otherwise identical verifier, so the advantage of the full system over direct web-search baselines should not be attributed solely to the database cascade.

\begin{table*}[htb]
\centering
\small
\renewcommand{\arraystretch}{1.14}
\setlength{\tabcolsep}{5pt}
\caption{Component ablation. Red arrows indicate the absolute drop in points relative to the full system.}
\label{tab:ablation}
\begin{tabularx}{\textwidth}{@{}>{\raggedright\arraybackslash}p{0.34\textwidth}YYYYY@{}}
\toprule
\textbf{Configuration} & \textbf{Exact $F_1$} & \textbf{Minor $F_1$} & \textbf{Major $F_1$} & \textbf{Macro $F_1$} & \textbf{Acc.} \\
\midrule
\rowcolor{green!10}
\textbf{\textsc{CiteCheck}} 
& \textbf{86.1} & \textbf{81.7} & \textbf{98.3} & \textbf{88.7} & \textbf{88.9} \\
\rowcolor{red!5}w/o Reviewer pass
& 84.1 \decrease{2.0} & 75.7 \decrease{6.0} & 93.3 \decrease{5.0} & 84.7 \decrease{4.0} & 84.9 \decrease{4.0} \\
\rowcolor{red!10}w/o Web search fallback
& 80.1 \decrease{6.0} & 72.7 \decrease{9.0} & 93.3 \decrease{5.0} & 81.7 \decrease{7.0} & 82.9 \decrease{6.0} \\
\bottomrule
\end{tabularx}
\end{table*}

\section{Retrieval Cascade Latency}
\label{app:cascade-latency}

We measure the latency of the retrieval cascade by running \citepCheck{}'s candidate-retrieval stage on all 982 citations in the benchmark after corruption injection. Each citation is evaluated once. The measurement includes the full retrieval cascade, including the web-search fallback when it is invoked, but excludes the downstream verifier and reviewer LLM calls. Records with API-side errors are excluded from aggregation.

Table~\ref{tab:cascade-latency-overall} reports overall latency. The full cascade takes 8.72 seconds per citation on average, with a median of 8.28 seconds. When excluding the web-search fallback, structured API retrieval alone takes 5.70 seconds on average and 4.02 seconds at the median. This indicates that web search improves coverage but contributes noticeably to latency.

\begin{table*}[t]
\centering
\small
\caption{Overall latency of the retrieval cascade over 982 citations.}
\label{tab:cascade-latency-overall}
\begin{tabularx}{\textwidth}{>{\raggedright\arraybackslash}Xrrrrrr}
\toprule
Setting & Mean & Median & P95 & P99 & Max & Std. \\
\midrule
Full cascade & 8.72s & 8.28s & 23.75s & 27.88s & 43.35s & 7.44s \\
Structured APIs only & 5.70s & 4.02s & 19.64s & 21.80s & 24.01s & 5.83s \\
\bottomrule
\end{tabularx}
\end{table*}

Table~\ref{tab:stage-latency} shows the per-stage latency and reach rate. Web search has the largest expected contribution to latency, adding 3.02 seconds per citation on average when counted over all citations. arXiv contributes 2.18 seconds and CrossRef contributes 1.76 seconds on average. Semantic Scholar is comparatively inexpensive, while OpenAlex has low median latency but a heavier tail.

\begin{table*}[t]
\centering
\small
\caption{Per-stage latency and reach rate of the retrieval cascade.}
\label{tab:stage-latency}
\begin{tabularx}{\textwidth}{>{\raggedright\arraybackslash}Xcccc}
\toprule
Stage & Reach & Cond. mean & Cond. median & Exp. contribution \\
\midrule
CrossRef & 80.5\% & 2.18s & 2.01s & 1.76s \\
Semantic Scholar & 56.4\% & 0.26s & 0.26s & 0.15s \\
OpenAlex & 55.7\% & 1.01s & 0.24s & 0.56s \\
arXiv & 68.9\% & 3.16s & 0.24s & 2.18s \\
Web search & 50.3\% & 6.00s & 5.40s & 3.02s \\
\bottomrule
\end{tabularx}
\end{table*}

Latency varies substantially by ground-truth class, as shown in Table~\ref{tab:cascade-latency-class}. Major hallucinations are the slowest, with a mean of 13.03 seconds, compared with 6.31 seconds for exact matches and 6.91 seconds for minor hallucinations. This is expected: exact and minor citations often resolve earlier in the cascade, while fabricated references are more likely to require later fallback stages or end in a not-found result.
\begin{table}[H]
\centering
\small
\caption{Retrieval cascade latency by ground-truth class.}
\label{tab:cascade-latency-class}
\begin{tabularx}{\columnwidth}{>{\raggedright\arraybackslash}Xrrrr}
\toprule
Class & Mean & Median & P95 & P99 \\
\midrule
\textsc{Exact} & 6.31s & 2.75s & 20.28s & 26.86s \\
\textsc{Minor} & 6.91s & 3.82s & 18.73s & 26.38s \\
\textsc{Major} & 13.03s & 10.60s & 25.51s & 29.99s \\
\bottomrule
\end{tabularx}
\end{table}

Finally, Table~\ref{tab:match-source-distribution} reports which stage produced the selected match. CrossRef produces the largest share of chosen matches, but web search and arXiv together account for a substantial fraction of the final candidates. This supports the use of a multi-source cascade rather than relying on a single scholarly index.

\begin{table}[H]
\centering
\small
\caption{Distribution of the retrieval stage that produced the chosen candidate.}
\label{tab:match-source-distribution}
\begin{tabularx}{\columnwidth}{>{\raggedright\arraybackslash}Xrr}
\toprule
Chosen source & Count & Share \\
\midrule
CrossRef & 278 & 28.3\% \\
Web search & 258 & 26.3\% \\
Not found & 195 & 19.9\% \\
arXiv & 191 & 19.5\% \\
OpenAlex & 53 & 5.4\% \\
Semantic Scholar & 7 & 0.7\% \\
\bottomrule
\end{tabularx}
\end{table}

\end{document}